\documentclass[aps,prb,fleqn,12pt,showpacs]{revtex4}
\usepackage{epsfig,color}
\usepackage{graphicx}
\usepackage{pdfpages}
\usepackage{braket}
\usepackage{amssymb,amsmath}
\usepackage{amsfonts} 
\usepackage{hyperref} 

\begin{document}
\title{Octahedral tilting induced isospin reorientation transition \\ in iridate heterostructures}
\author{Shubhajyoti Mohapatra}
\author{Sreemayee Aditya}
\author{Rohit Mukherjee}
\author{Avinash Singh}
\email{avinas@iitk.ac.in}
\affiliation{Department of Physics, Indian Institute of Technology, Kanpur - 208016, India}
\date{\today} 
\begin{abstract}  
Iridate heterostructures are gaining interest as their magnetic properties are much more sensitive to structural distortion compared to pure spin systems due to spin-orbital entanglement induced by strong spin-orbit coupling. While bulk monolayer and bilayer iridates show $ab$-plane canted and $c$-axis antiferromagnetic (AFM) order, recent experiments on layered iridate superlattices (SL) have revealed striking properties, especially in the bilayer SL. A spin model is presented including the tilting induced Kitaev type interactions, which illustrates the proclivity towards $ab$-plane canted AFM order. A realistic Hubbard model including spin-dependent hopping terms arising from octahedral rotation and tilting is constructed for the bilayer SL in isospin space, and magnetic excitations are investigated in the self-consistently determined magnetic state. The Hubbard model analysis confirms the spin model results and shows strongly reduced magnon energy gap and an isospin reorientation transition from $c$-axis to $ab$-plane canted AFM order with increasing tilting. 

\end{abstract}
\maketitle

\newpage
\section{Introduction}


Strongly spin-orbit-coupled systems are promising candidates for artificial heterostructures with leveraged magnetic properties arising from the sensitive coupling of magnetic moments to structural distortion.\cite{hwang_Nature_2012,chakhalian_RMP_2014,hao_JPCS_2019} Recently, layered iridate superlattices nSIO/1STO (with n=1,2,3) have been synthesized by stacking alternating layers of $\rm n SrIrO_3$ and $\rm SrTiO_3$.\cite{matsuno_PRL_2015,sykim_PRB_2016,hao_PRL_2017,mayers_PRL_2018,meyers_SREP_2019} Investigations on these heterostructures have highlighted structural distortion effects on  magnetic order, magnon energy gap, and magnetic order switching. 

The monolayer and bilayer superlattices (nSIO/1STO, n=1,2) display canted antiferromagnetic (AFM) order ($ab$-plane for n=1, $c$-axis for n=2 with ferromagnetic moment in the $ab$ plane).\cite{meyers_SREP_2019} Although for 1SIO/STO, this behaviour is  similar to the bulk monolayer iridate, the significant ferromagnetic moment measured in 2SIO/1STO is attributed to the presence of octahedral tilting which induces canting of the $c$-axis AFM moments. Study of magnetic excitations by resonant inelastic x-ray scattering (RIXS) shows that both superlattices have similar magnon dispersions within experimental resolution.\cite{meyers_SREP_2019} Both samples show finite magnon gap at $(\pi,\pi)$. While the magnon gap ($\approx 32$ meV) for 1SIO/1STO is nearly same as for $\rm Sr_2 IrO_4$, the measured gap ($\approx 57$ meV) for 2SIO/1STO is significantly reduced compared to the gap ($\approx 90$ meV) for the bilayer bulk compound $\rm Sr_3 Ir_2 O_7$.\cite{sala_PRB_2015} This large reduction in magnon energy gap indicates emergence of new magnetic interactions due to structural distortions. 

Octahedral tilting was identified as the important structural feature which distinguishes the bilayer superlattice from the bulk compound. While both bulk monolayer and bilayer iridates feature large in-plane octahedral rotations, there is no octahedral tilting in $\rm Sr_2 IrO_4$ and very small tilting in $\rm Sr_3 Ir_2 O_7$.\cite{hogan_PRB_2016} However, 2SIO/1STO shows large octahedral tilting, although this effect is negligible in 1SIO/1STO.\cite{matsuno_PRL_2015,bkim2_PRB_2017,meyers_SREP_2019} Recent pressure studies on the bulk bilayer iridate also show presence of octahedral tilting and magnon softening, supporting the key role of structural distortion.\cite{zhang_NPJ_2019} Significant octahedral tilting was suggested in the above studies to be responsible for driving the bilayer bulk and superlattice (SL) systems from $c$-axis towards $ab$-plane AFM ordering. 

Detailed theoretical investigations explicitly including the additional anisotropic interaction terms generated by the octahedral tilting in the SL have not been carried out. Earlier studies based on simplistic spin models have considered the same type of anisotropic interaction terms as for the bulk bilayer iridate, and have extracted changes in the SL by fitting with the magnon dispersion.\cite{meyers_SREP_2019} In this work, we will therefore investigate the octahedral tilting induced proclivity towards the $ab$-plane canted AFM order as well as the magnon gap reduction using a realistic bilayer Hubbard model including appropriate spin-dependent hopping terms corresponding to octahedral rotation and tilting in the bilayer SL. The spin-dependent hopping terms incorporate the additional orbital mixings between the $xy$ and the $xz,yz$ orbitals induced by octahedral tilting. 

The structure of this paper is as below. After briefly reviewing the microscopic origin of the additional anisotropic interaction terms for the bilayer SL in Sec. II, a minimal spin model is presented in Sec. III which reveals a reorientation transition from $c$-axis to $ab$-plane AFM order with increasing octahedral tilting. A realistic bilayer Hubbard model including appropriate spin-dependent hopping terms corresponding to octahedral rotation and tilting in the bilayer SL is introduced in Sec. IV. Electronic band structure, self-consistent determination of magnetic order, and magnon excitations are discussed in Secs. IV and V. Finaly, some key conclusions are presented in Sec. VI. 

\section{Octahedral tilting and anisotropic interactions} 

While $c$-axis AFM order is stabilized in the bulk bilayer iridate, the 2SIO/1STO heterostructure is on the verge of a reorientation transition to $ab$-plane canted AFM order. This  reorientation transition is driven by the octahedral tilting, and the microscopic origin of the relevant anisotropic spin interactions is briefly discussed below. 

Due to strong SOC, the $t_{2g}$ manifold is split into the effective $J=1/2$ and 3/2 sectors, of which only the half-filled $J=1/2$ sector is magnetically active. The two isospin states for the $J=1/2$ sector are given by:
\begin{eqnarray}
|\tau = \uparrow \rangle &=& (1/\sqrt{3})[|yz,\downarrow\rangle + i|xz,\downarrow\rangle + |xy,\uparrow\rangle ] \nonumber \\
|\tau = \downarrow \rangle &=& (1/\sqrt{3})[|yz,\uparrow\rangle - i|xz,\uparrow\rangle - |xy,\downarrow\rangle ] 
\label{soe}
\end{eqnarray}
in terms of the local orbital-spin basis states $|\mu,\sigma\rangle$, where $\mu=yz,xz,xy$ and $\sigma = \uparrow,\downarrow$. The interplay of these SOC-induced spin-orbital-entangled states and the orbital mixing hopping terms generated by octahedral rotation and tilting is essentially responsible for the anisotropic magnetic interactions. 

We first consider the staggered octahedral rotation (small angle $\alpha$) about the $c$ axis, which generates orbital mixing hopping terms $t_{ij}^{yz|xz} = -t_{ij}^{xz|yz}$ between neighboring sites $i$ and $j$, where $t_{ij}^{yz|xz} \approx t_\pi \sin \alpha \cos \alpha \approx t_\pi \sin \alpha$, corresponding to $\pi$ overlap between $yz$ and $xz$ orbitals. Using the above transformation, the hopping Hamiltonian in the $J=1/2$ sector:
\begin{equation}
T=-\sum_{\langle ij \rangle,\tau\tau'} |i\tau \rangle \left [ t {\bf 1} + i \sigma_z t_z \right ]_{\tau\tau'} \langle j\tau' | \; ,
\label{hopping}
\end{equation}
where the usual spin-independent hopping term:
\begin{equation}
t = \frac{1}{3} \left (t_{ij}^{xz|xz} + t_{ij}^{yz|yz} + t_{ij}^{xy|xy} \right ) = \frac{1}{3} \left ( 2 t_\pi + t_\delta \right )
\end{equation}
involving the $\pi$ and $\delta$ overlaps of the $yz,xz,xy$ orbitals, and the spin-dependent hopping term (with the Pauli matrix $\sigma_z$) arises from the orbital mixing hopping terms:
\begin{equation}
t_z = \frac{1}{3} \left (t_{ij}^{yz|xz} - t_{ij}^{xz|yz} \right ) \approx \frac{2}{3} t_\pi \sin \alpha .
\label{orb_mix_hop}
\end{equation}

Including the local interaction term $Un_{i\uparrow} n_{i\downarrow}$, and carrying out the usual strong-coupling expansion for the half-filled Hubbard model up to second order in the hopping terms, yields the Kitaev-type $K_z (S_i^z S_j^z - S_i^x S_j^x - S_i^y S_j^y)$ and Dzyaloshinski-Moriya (DM) $2\sqrt{J K_z} \hat{z}.({\bf S}_i \times {\bf S}_j)$ interactions besides the usual isotropic Heisenberg interaction $J {\bf S}_i . {\bf S}_j$, where $J=4t^2/U$ and $K_z=4t_z^2/U$.\cite{honeycomb_JMMM_2019}

Similarly, octahedral tilting about the $a$ and $b$ crystal axes will generate orbital mixing hopping terms between the $xy$ and the $xz/yz$ orbitals, leading to spin-dependent hopping terms $t_x$ and $t_y$, respectively. The hopping Hamiltonian in Eq. (\ref{hopping}) will then include $i \sigma_x t_x$ and $i \sigma_y t_y$ contributions. Besides additional Kitaev-type and DM interactions, the $t_x$ and $t_y$  terms will generate symmetric-off-diagonal (SOD) interaction terms $\Gamma_{\alpha\beta} (S_i^\alpha S_j^\beta + S_i^\beta S_j^\alpha)$, where $\Gamma_{\alpha\beta}=8t_\alpha t_\beta/U$ and $\alpha,\beta = x,y,z$. In the next section, these additional anisotropic interaction terms will be shown to be responsible for a octahedral tilting driven isospin reorientation transition.

\section{Spin model}

A minimal spin model is presented here which illustrates the proclivity towards $ab$-plane canted AFM order and the reduced magnon energy gap in the 2SIO/STO superlattice compared to the bulk bilayer iridate which shows robust $c$-axis AFM order. This model incorporates the critical role of the tilting induced Kitaev type interactions and reveals a reorientation transition from $c$-axis to $ab$-plane canted AFM order with increasing octahedral tilting.

In the bulk monolayer and bilayer compounds, the octahedral tilting is negligible, and the three-orbital model therefore features only the orbital mixing hopping terms between the $xz$ and $yz$ orbitals arising from the staggered octahedral rotations about the $c$ axis.\cite{carter_PRB_2013,iridate1_PRB_2017} This orbital mixing generates a spin-dependent hopping term, which results in anisotropic magnetic interactions in the strong coupling expansion. While the spin-dependent hopping term can be gauged away for the monolayer case (hence no true magnetic anisotropy), $c$-axis ordering with a large magnon gap is obtained in the bilayer case due to a frustration effect involving different canting proclivities for in-plane and out-of-plane neighboring spins.\cite{iridate1_PRB_2017}

In contrast, the 2SIO/STO superlattice is characterized by both octahedral rotation as well as tilting which are comparable in magnitude, resulting in additional orbital mixing hopping terms between the $xy$ orbital and the $xz,yz$ orbitals. This leads to additional anisotropic magnetic interactions in the isospin ($J=1/2$) model corresponding to the spin-dependent hopping terms $t_{x,y,z}$. We consider a minimal model: 
\begin{eqnarray}
H_{\rm eff} & = & \sum_{\langle ij \rangle} \left [ J {\bf S}_i . {\bf S}_j + \left(K_c+K_z-K_x-K_y\right)S_i^z S_j^z \right . \nonumber \\
& + & \left(K_x-K_y-K_z\right)S_i^x S_j^x + \left(K_y-K_z-K_x\right)S_i^y S_j^y \nonumber \\
& + & \left . 2\sqrt{JK_{z}}\left(S_i^xS_j^y - S_j^x S_i^y \right) \right ]
\label{model}
\end{eqnarray}
where $K_c$ incorporates the effective $c$-axis anisotropy arising from the frustration effect in the bilayer compound. The last term is the DM interaction which is responsible for the $ab$-plane canting. We first consider several limiting cases in order to connect to the bulk monolayer and bilayer iridate compounds. 

(1) $K_x=K_y=K_c=0$: this case corresponds to the bulk monolayer iridate. For $c$-axis AFM order with $S_i ^z=S$ and $S_j ^z=-S$ on the two sublattice sites, the classical energy:
\begin{align}
E^c = \sum_{\langle ij \rangle} -\left(J+K_z\right) S^2.
\end{align}
On the other hand, for $ab$-plane canted AFM order, with $S_i ^x=-S_j ^x=-S\cos\phi$ and $S_i ^y=S_j ^y=S\sin\phi$ corresponding to canting angle $\phi$, we obtain: 
\begin{align}
E^{ab} (\phi) = \sum_{\langle ij \rangle}\left(-J\cos2\phi+K_z\cos2\phi-2\sqrt{JK_z}\sin2\phi\right) S^{2},
\end{align}
minimizing which with respect to $\phi$ yields:
\begin{align}
\tan2\phi = \frac{2\sqrt{JK_z}}{J-K_z} = \frac{2\sqrt{K_z/J}}{1-K_z/J} = \frac{2\tan\phi}{1-\tan^{2}\phi}.
\end{align}
With the optimal canting angle given by $\tan\phi ^* =\sqrt{\frac{K_z}{J}} = \frac{t_z}{t}$, the minimum energy $E^{ab}(\phi^*)=E^c$, identical to the energy for $c$-axis AFM order. This degeneracy reflects the absence of true magnetic anisotropy and is responsible for the nearly gapless magnon mode in the bulk monolayer iridate compound.

(2) $K_c>0$, $K_x=K_y=0$: this case corresponds to the bulk bilayer iridate. The extra energy gain for $c$-axis order in this case breaks the degeneracy, resulting in true magnetic anisotropy and the large magnon gap in the bulk bilayer iridate.

(3) $K_c>0$, $K_x > 0$, $K_y = 0$: this case corresponds to the bilayer iridate superlattice. When $K_x$ becomes finite (due to octahedral tilting), the extra energy gain for $c$-axis AFM order is reduced to $\left(K_c-K_x\right)$, while there is an extra energy gain for $ab$-plane canted AFM order. This suggests that with increasing $K_x$, there must be a reorientation transition from $c$-axis to $ab$-plane canted AFM order.

For $c$-axis AFM order (with $S_i=S$ and $S_j=-S$), we obtain from Eq. (\ref{model}): 
\begin{align}
E^c = -\sum_{\langle ij \rangle }\left(J+K_{z}+K_c-K_{x}\right) S^2,
\end{align}
whereas for $ab$-plane canted AFM order (with $S_i^x$=$-S_j^x$=$-S\cos\phi$ and $S_i^y$=$S_j^y$=$S\sin\phi$): 
\begin{align}
E^{ab} (\phi)=\sum_{\langle ij \rangle}\left(\left(-J+K_{z}\right) \cos 2\phi -K_{x}-2\sqrt{JK_{z}}\sin2\phi\right) S^{2}. 
\end{align}
Minimization yields the same condition $\tan\phi^*=\sqrt{\frac{K_z}{J}}$ for the optimal canting angle $\phi^*$, and we obtain for the minimum energy:
\begin{align}
E^{ab} (\phi ^*) =-\sum_{\langle ij \rangle} \left(J+K_{z}+K_{x}\right) S^{2}.
\end{align}
For $K_x=0$, we have $E^c < E^{ab} (\phi^*)$, confirming the true magnetic anisotropy as in case (2). However, with increasing $K_x$, the energy difference decreases, and the $ab$-plane canted AFM order becomes the ground state for $K_x > K_x^*$, where the critical value for the transition:
\begin{equation}
K_x^* = K_c/2
\end{equation}
is simply related to the frustration-induced $c$-axis anisotropy term $K_c$.   

(4) $K_c>0$, $K_x = K_y > 0$: this is a more realistic case for the bilayer iridate superlattice accounting for octahedral tiltings around both $a$ and $b$ crystal axes. Carrying out a general analysis here for arbitrary polar angle $\theta$, with $S_i^x=-S_j^y=-S\sin\theta\sin\phi$, $S_i^y=-S_j^x=-S\sin\theta\cos\phi$, and $S_i^z=-S_j^z=S\cos\theta$ for the A and B sublattice sites. This order accounts for the expected planar AFM order along the rotated $\hat{a'}=(\hat{a}+\hat{b})/\sqrt{2}$ direction and canting along the normal $\hat{b'}=(-\hat{a}+\hat{b})/\sqrt{2}$ direction. We obtain for the classical energy:
\begin{eqnarray}
E(\theta,\phi) &=& \sum_{\langle ij \rangle} \left [-(J+K_z+K_c-2K_x)\cos^2 \theta \right . \nonumber \\
&-& \left . \left ( (J-K_z)\cos2\phi' + 2\sqrt{JK_z}\sin2\phi' \right )\sin^2 \theta 
\right ] S^{2},
\end{eqnarray}
where $2\phi=\pi/2 + 2\phi'$. With the optimal canting angle given by $\tan\phi'^*=\sqrt{\frac{K_z}{J}}$, we obtain: 
\begin{equation}
E(\theta,\phi^*) = \sum_{\langle ij \rangle} \left [-(J+K_z) -(K_c-2K_x)\cos^2 \theta \right ] S^{2} ,
\end{equation}
which has energy minimum at $\theta^*$=0 ($c$-axis order) for $K_x < K_c/2$ and at $\theta^*$=$\pi/2$ ($ab$-plane order) for $K_x > K_c/2$. The octahedral tilting therefore reduces the effective $c$-axis anisotropy term to $K_c-2K_x$ in the superlattice. The spin model analysis presented here provides a minimal realization of the octahedral tilting driven reorientation transition at the critical value $K_x^*=K_y^*= K_c/2$. The realistic bilayer Hubbard model analysis discussed in the next section shows that the bilayer iridate SL may be quite close to this transition point. 

Effects of the other anisotropic interaction terms associated with the octahedral tilting on the magnetic order is qualitatively discussed below. The Kitaev interactions generated by $t_x$ and $t_y$ are fully included above. The DM interactions $D_{x,y}(\hat{x} + \hat{y}).({\bf S}_i \times {\bf S}_j)$ generated by $t_x$ and $t_y$ will generally induce spin canting. The SOD terms $-\Gamma_{xz} (S_i^x S_j^z + S_i^z S_j^x)$ and $-\Gamma_{yz} (S_i^y S_j^z + S_i^z S_j^y)$ (generated by the products $t_x t_z$ and $t_y t_z$) will contribute to AFM order in the $x$ and $y$ directions. These effects will be quantitatively investigated below within a realistic bilayer Hubbard model for the superlattice including spin-dependent hopping terms associated with octahedral rotation and tilting. 

\begin{figure}
\vspace*{-20mm}
\hspace*{-15mm}
\psfig{figure=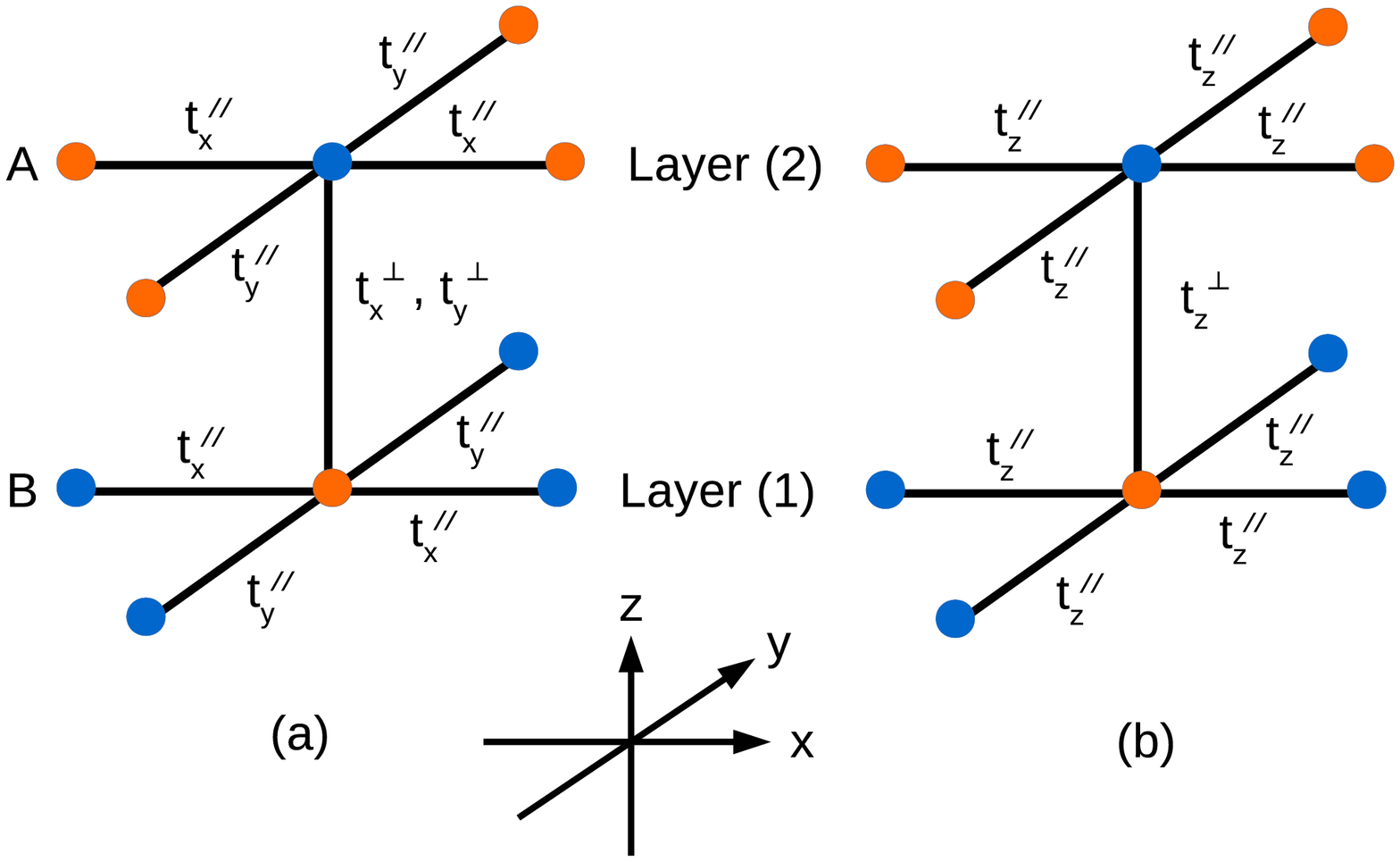,angle=0,width=90mm}\hspace{-3mm}
\psfig{figure=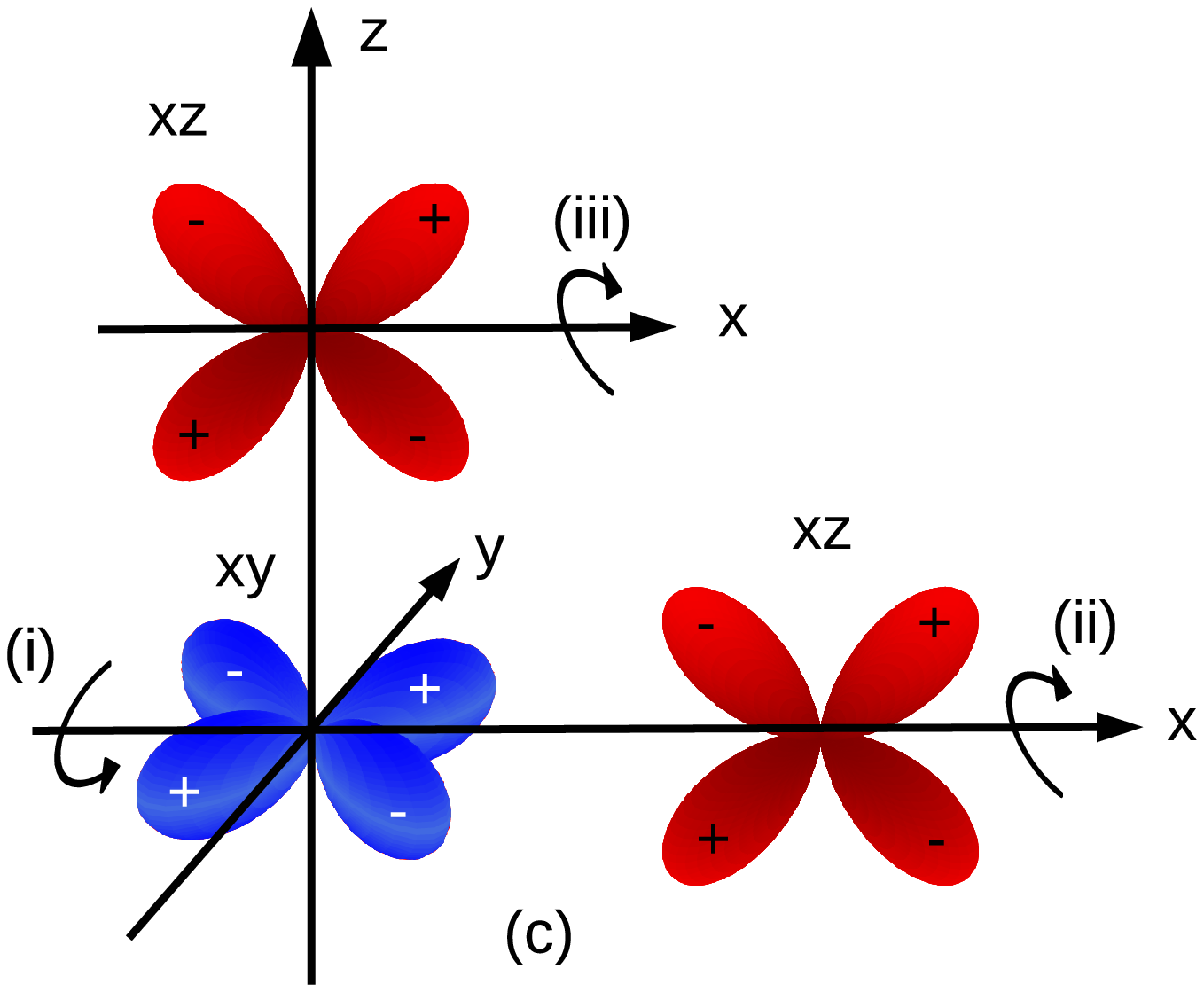,angle=0,width=90mm}\hspace{-25mm}
\vspace{-35mm}
\caption{The $x,y$ components (a) and $z$ component (b) of the spin-dependent hopping terms for intra- and inter-layer neighboring sites (indicated by $\parallel$ and $\perp$). Here A and B denote the two magnetic sublattices. The spin-dependent hopping terms are antisymmetric (${\bf t'_{ji}}=-{\bf t'_{ij}}$) due to the staggered octahderal rotation and tilting. (c) shows $xy$-$xz$ orbital mixing ($\pi$ overlap) due to both tiltings (i) and (ii) for intra-layer neighbors, but due to only (i) for inter-layer neighbors.} 
\label{hopping_terms}
\end{figure}

\section{Bilayer Hubbard model}

For the 2SIO/1STO SL, we consider a minimal bilayer Hubbard model:
\begin{equation}
H = -\sum_{\langle ij \rangle,\sigma\sigma'} 
\left ( t_{ij} {\bf 1} + i \makebox{\boldmath $\sigma . t'_{ij}$} \right )_{\sigma\sigma'} 
a_{i\sigma} ^\dagger a_{j\sigma'} + H.c. 
+ U \sum_i n_{i\uparrow} n_{i\downarrow}
\label{bilayer_Hubbard}
\end{equation}
on a square lattice for each layer, where the sum $\langle ij \rangle$ includes intra-layer and inter-layer pairs of lattice sites $i,j$. Here $t_{ij}$ and ${\bf t'_{ij}}$ are the spin-independent and spin-dependent hopping terms, respectively. For $t_{ij}$, we have included first, second, and third neighbor intra-layer hopping terms $(t_1 ^\parallel, t_2 ^\parallel, t_3 ^\parallel)$ and first neighbor inter-layer hopping term ($t_1 ^\perp$). Only the first neighbor spin-dependent hopping terms ${\bf t'_{ij}} = (t_x,t_y,t_z)$ are included, as shown in Fig. 1 for the intra- and inter-layer sites.  

In the Hartree-Fock (HF) approximation, the interaction term reduces to a local exchange-field interaction:  
\begin{equation}
\mathcal{H}_{\rm int} ^{\rm HF} = 
\sum_{{\bf k} s}  \psi_{{\bf k}s}^{\dagger} 
\begin{pmatrix} - \makebox{\boldmath $\sigma . \Delta^s$}
\end{pmatrix} \psi_{{\bf k}s} 
= - \sum_{{\bf k} s} \psi_{{\bf k}s}^{\dagger} 
\begin{pmatrix} \Delta_z ^s & \Delta_x ^s - i \Delta_y ^s \\
\Delta_x ^s + i \Delta_y ^s & -\Delta_z ^s  \\
\end{pmatrix} \psi_{{\bf k}s}
\label{gen_ord_dirn} 
\end{equation}  
where $\psi_{{\bf k}s} ^\dagger = (a_{{\bf k}s\uparrow} ^\dagger \; \; a_{{\bf k}s\downarrow} ^\dagger)$, $s$ is the composite layer-sublattice index corresponding to the two layers (1,2) and the two sublattices (A/B), and the exchange field components $\Delta_{x,y,z} ^s$ are self-consistently determined from:
\begin{equation}
2 \Delta_{x,y,z} ^s = U m_{x,y,z} ^s 
\label{selfcon}
\end{equation}
in terms of the magnetization components. 

We consider a composite 2-layer$\otimes$2-sublattice$\otimes$2-spin basis to represent the HF Hamiltonian matrix with appropriate hopping terms in the ${\bf k}=(k_x,k_y)$ space. For the self-consistent determination of the exchange field components, an iterative approach was employed starting with an initial choice for $(\Delta^s_x,\Delta^s_y,\Delta^s_z)$ with staggered order on the two layers and sublattices. In each iteration step, the local magnetization components were evaluated using the eigenvectors and eigenvalues of the Hamiltonian matrix, and the exchange field components were updated using Eq. (\ref{selfcon}). 

\begin{figure}
\vspace*{0mm}
\hspace*{0mm}
\psfig{figure=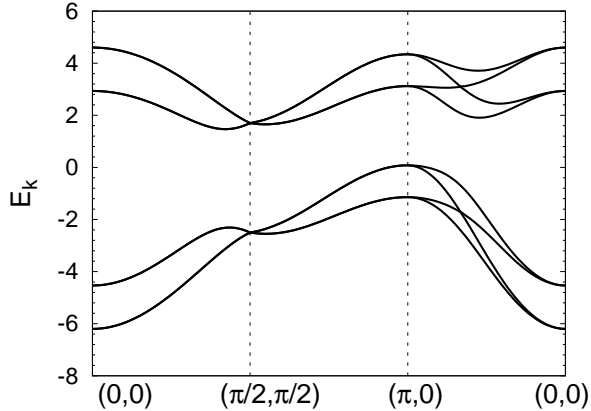,angle=0,width=80mm}
\vspace{0mm}
\caption{Calculated electronic band structure in the self-consistent AFM state of the bilayer Hubbard model with  hopping parameter values as given in Table \ref{table-1}.} 
\label{band_struc}
\end{figure}

The calculated electronic band structure in the self-consistent state is shown in Fig. 2 for the bilayer hopping parameter values given in Table I. The Coulomb interaction value $U = 5.2 |t_1| \sim 0.8$ eV considered above is same as for the bulk bilayer compound.\cite{iridate1_PRB_2017} The calculated band structure is in qualitative agreement with DFT study for the 2SIO/1STO superlattice,\cite{sykim_PRB_2016} including features such as valence band maximum at $(\pi,0)$, conduction band minimum near $(\pi/2,\pi/2)$, overall bandwidth, and the band splitting structure.  

The magnetization components obtained are: $(m_x ^A, m_y ^A, m_z ^A)=(0.04,-0.119,0.697)$ and $(m_x ^B, m_y ^B, m_z ^B)=(0.119,-0.04,-0.697)$, indicating dominantly $c$-axis AFM order. Furthermore, finite ferromagnetic moment oriented along the $a$-$b$ diagonal is evident from $(m_x ^A + m_x ^B)$=$-(m_y ^A + m_y ^B)$$\ne$0 and $(m_z ^A + m_z ^B)$=0. The bilayer magnetic order therefore corresponds to dominantly $c$-axis AFM order with $ab$-plane canting. This magnetic order will be referred to as phase I. The overall magnetic order in this phase is described by antiferromagnetic ${\bf m}_{\rm AF} \equiv ({\bf m}_{\rm A} - {\bf m}_{\rm B})/2 = m_{\rm AF}^{\rm c} \hat{z} - m_{\rm AF}^{\rm ab} (\hat{x} + \hat{y})$ and ferromagnetic ${\bf m}_{\rm F} \equiv ({\bf m}_{\rm A} + {\bf m}_{\rm B})/2 = m_{\rm F}^{\rm ab} (\hat{x} - \hat{y})$ components, which yields:
\begin{eqnarray}
{\bf m}_{\rm A} & = & + m_{\rm AF}^{\rm c} \hat{z} - (m_{\rm AF}^{\rm ab} - m_{\rm F}^{\rm ab}) \hat{x}  - (m_{\rm AF}^{\rm ab} + m_{\rm F}^{\rm ab}) \hat{y} \nonumber \\
{\bf m}_{\rm B} & = & - m_{\rm AF}^{\rm c} \hat{z} + (m_{\rm AF}^{\rm ab} + m_{\rm F}^{\rm ab}) \hat{x} + (m_{\rm AF}^{\rm ab} - m_{\rm F}^{\rm ab}) \hat{y} .
\end{eqnarray}

With increasing $t_{x,y}^{\parallel}$ and keeping $t_{x,y}^{\perp}=t_{x,y}^{\parallel}/2$, we find an isospin reorientation transition at a critical value ($\approx 0.25$) to a dominantly $ab$-plane canted AFM order which will be referred to as phase II. Near the critical value, self consistency required several thousand iterations. For $t_{x,y}^{\parallel}=0.3$ (in phase II), the magnetization components obtained are: $(0.333,-0.594,-0.144)$ and $(-0.594,0.333,-0.144)$ for the A and B sublattices, respectively. We have kept $t_{x,y}^{\perp}=t_{x,y}^{\parallel}/2$ to account for the nominally double orbital mixing hopping terms ($\pi$ overlap) between $xy$ and $xz,yz$ orbitals for in-plane versus out-of-plane neighbors [Fig. 1(c)]. The behavior of magnon excitations through the reorientation transition will be discussed in the next section. 

\begin{table}[!]
\centering
\caption{Bilayer hopping parameter values in terms of the energy scale unit $|t_{1}^\parallel|$ (=150 meV).} 
\label{table-1} \vspace{5mm}
\begin{ruledtabular}
\begin{tabular}{c  c  c  c  c  c  c  c}
$t_1 ^\parallel$ & $t_2 ^\parallel$ & $t_3 ^\parallel$ & $t_{x}^{\parallel}=t_{y}^{\parallel}$ & $t_{z}^{\parallel}$ & $t_{1}^{\perp}$ & $t_{x}^{\perp}=t_{y}^{\perp}$ & $t_{z}^{\perp}$ \\ 
\hline 
-1.0 & 0.3 & -0.1 & 0.25 & -0.2 & -0.8 & 0.125 & -0.6\\
\end{tabular} 
\end{ruledtabular}
\end{table}

\section{Magnon excitations}

In the following, we will investigate transverse spin fluctuations in the self-consistent magnetic state obtained above. We therefore consider the time-ordered magnon propagator: 
\begin{equation}
\chi ({\bf q},\omega) = \int dt \sum_{i} e^{i\omega(t-t^\prime)}
e^{-i{\bf q}.({\bf r}_i - {\bf r}_j)}  
\langle \Psi_0 | T [ S_{i} ^{\mu} (t) S_{j} ^{\nu} (t^\prime) ] | \Psi_0 \rangle
\label{chi}
\end{equation}
involving the $\mu,\nu=x,y,z$ components of the isospin operators $S_{i}^{\mu}$ and $S_{j}^{\nu}$ at lattice sites $i$ and $j$. In the random phase approximation (RPA), the magnon propagator is obtained as:
\begin{equation}
[\chi({\bf q},\omega)] = \frac{[\chi^0({\bf q},\omega)]}
{1 - 2U] [\chi^0({\bf q},\omega)]}
\label{eq:spin_prop}  
\end{equation}
where the bare particle-hole propagator:
\begin{equation}
[\chi^0 ({\bf q},\omega)]_{ss'} ^{\mu \nu} = \frac{1}{4} \sum_{{\bf k}} \left [ 
\frac{ 
\langle \varphi_{\bf k-q} | \sigma^\mu | \varphi_{\bf k} \rangle_s
\langle \varphi_{\bf k} | \sigma^\nu | \varphi_{\bf k-q} \rangle_{s'} 
} 
{E^+_{\bf k-q} - E^-_{\bf k} + \omega - i \eta }
+ \frac{
\langle \varphi_{\bf k-q} | \sigma^\mu | \varphi_{\bf k} \rangle_s
\langle \varphi_{\bf k} | \sigma^\nu | \varphi_{\bf k-q} \rangle_{s'} 
} 
{E^+_{\bf k} - E^-_{\bf k-q} - \omega - i \eta } \right ]
\end{equation}
was evaluated in the composite spin-sublattice-layer basis (3 spin components $\otimes$ 2 sublattices $\otimes$ 2 layers) by integrating out the fermions in the self-consistent ground state. Here $E_{\bf k}$ and $\varphi_{\bf k}$ are the eigenvalues and eigenvectors of the Hamiltonian matrix, the indices $s,s'=1,4$ correspond to the layer-sublattice subspace, and the superscript $+(-)$ refers to particle (hole) energies above (below) the Fermi energy. 

\begin{figure}
\vspace*{0mm}
\hspace*{0mm}
\psfig{figure=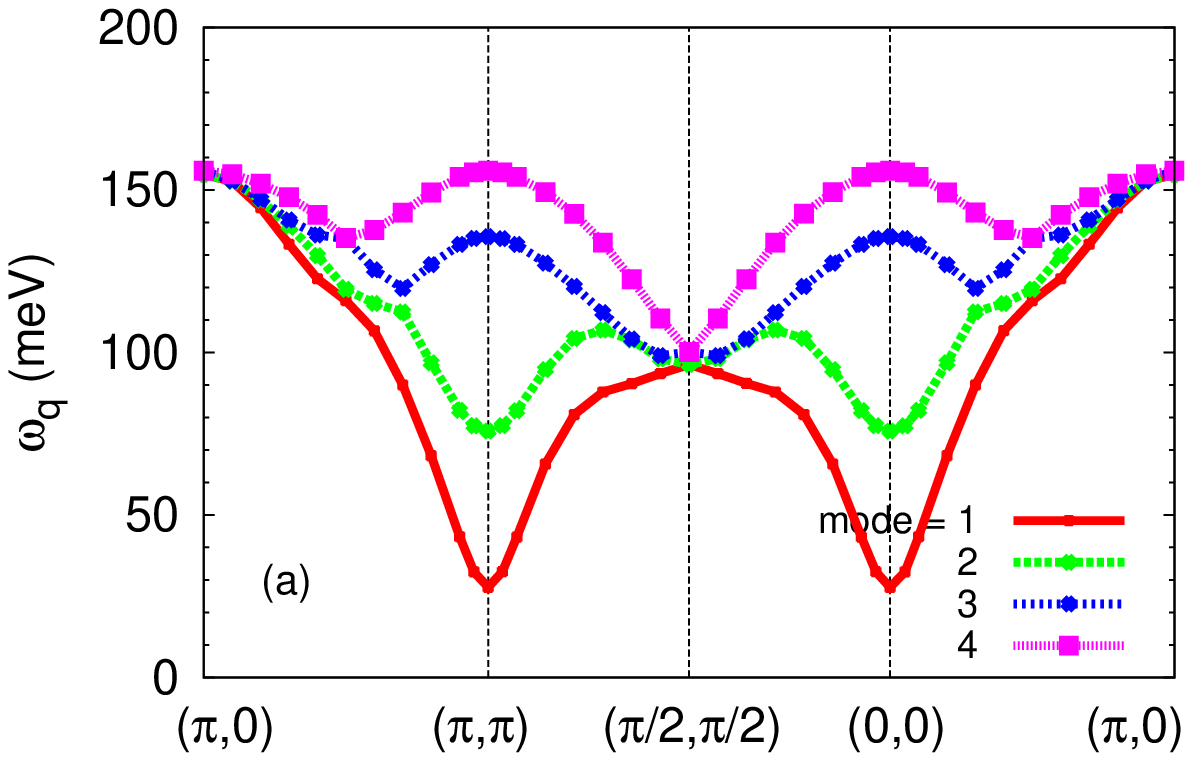,angle=0,width=80mm}
\psfig{figure=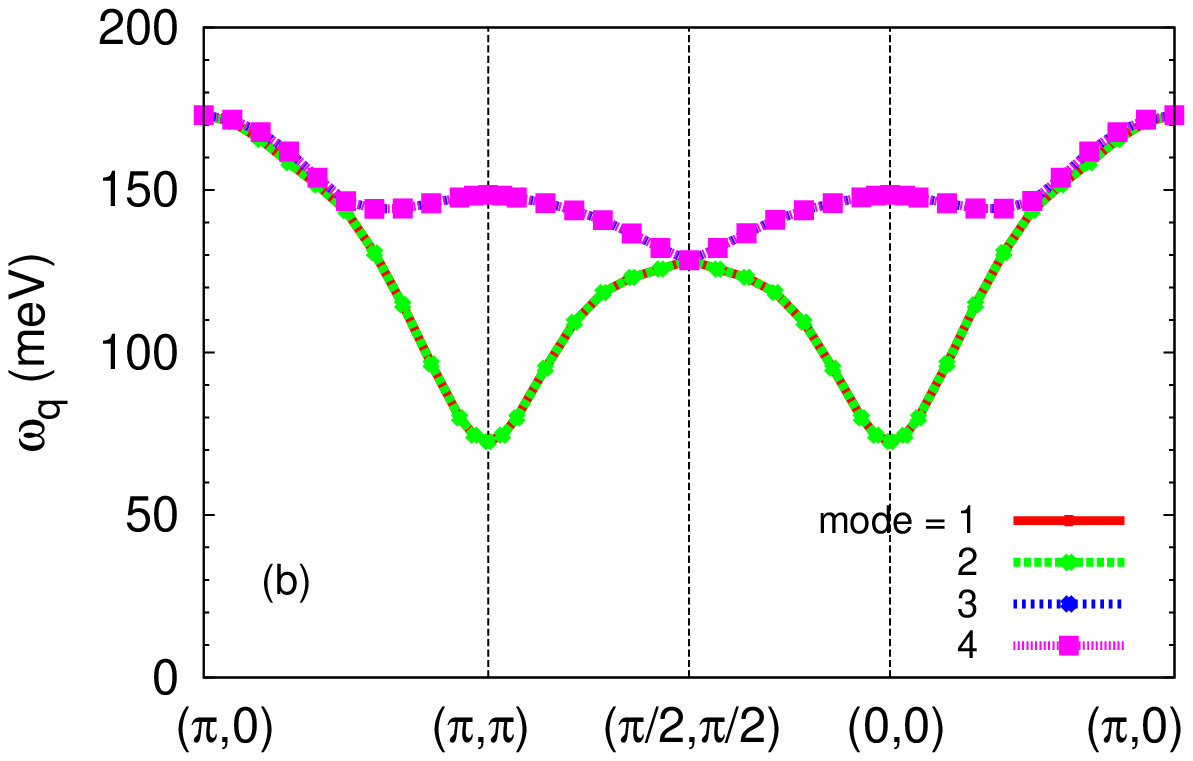,angle=0,width=80mm}
\vspace{0mm}
\caption{Calculated magnon dispersions for the bilayer Hubbard model (a) with and (b) without the hopping terms $t_x,t_y$, showing strong magnon gap reduction (from 75 meV to 30 meV) induced by the octahedral tilting in the bilayer SL. Here, the bilayer reference case (b) corresponds to reduced octahedral rotation effect but no tilting.} 
\label{magnon}
\end{figure}

In the following, we will focus on the magnon energies $\omega_{\bf q}$ obtained from Eq. \ref{eq:spin_prop} using the pole condition $1-2U \lambda_{\bf q}(\omega)=0$, where $\lambda_{\bf q} (\omega)$ is the eigenvalue of the $[\chi^0 ({\bf q},\omega)]$ matrix. The $12 \times 12$  $[\chi^0 ({\bf q},\omega)]$ matrix was evaluated by performing the $\bf k$ sum over the 2D Brillouin zone divided into a 300 $\times$ 300 mesh. Using this approach, magnon excitations were studied earlier for the bilayer bulk compound $\rm Sr_3 Ir_2 O_7$.\cite{iridate1_PRB_2017} Two modes corresponding to acoustic and optical branches were obtained, with the acoustic mode showing a large magnon gap arising from the frustration effect due to different canting proclivities of in-plane and out-of-plane neighboring spins. 

The calculated magnon dispersions for the bilayer Hubbard model is shown in Fig. \ref{magnon}, and compared for two cases corresponding to (a) the bilayer SL and (b) the bilayer reference. For the SL case, same hopping parameters were used as for the band structure study (Table I). For the reference case, the hopping terms $t_{x,y}^{\parallel}$ and $t_{x,y}^{\perp}$ were set to zero as the octahedral tilting is negligible. All other hopping terms were kept same for simplicity. The reference case corresponds to reduced octahedral rotation compared to the bulk case. Correspondingly, the magnon gap is reduced to $\approx 75$ meV [Fig. 3(b)] in the reference case (with $t_z^\perp = -0.6$) compared to $\approx 90$ meV in the bulk case (with $t_z^\perp = -0.7$).   

Comparison of the calculated magnon dispersions therefore exclusively highlights the role of spin-dependent hopping terms $t_x,t_y$ associated with octahedral tilting in the bilayer SL. The most important effect as seen in Fig. \ref{magnon} is the strong magnon gap reduction from $\sim 75$ meV to $\sim 30$ meV in the bilayer SL. Also, the two-fold degeneracy in the bulk case is lifted, and the acoustic and optical modes are further split due to the spin mixing terms $t_{x,y}$. 

Furthermore, all four branches in Fig. 3(a) are degenerate at $(\pi, 0)$ and $(\pi/2, \pi/2)$, where the energies are $\sim$ 150 and 100 meV, respectively. The significant $(\pi,0)$ magnon energy reduction from 170 meV in the bulk case to 150 meV in the SL case is due to minute enhancement in $t_2 ^\parallel$, which follows from the reduced octahedral rotation and therefore enhanced $xy$ orbital overlap. These features are in excellent agreement with the RIXS spectra of 2SIO/1STO superlattice.\cite{meyers_SREP_2019} Due to experimental limitations, the higher-energy magnon modes have not yet been experimentally resolved. 

\begin{figure}
\vspace*{0mm}
\hspace*{0mm}
\psfig{figure=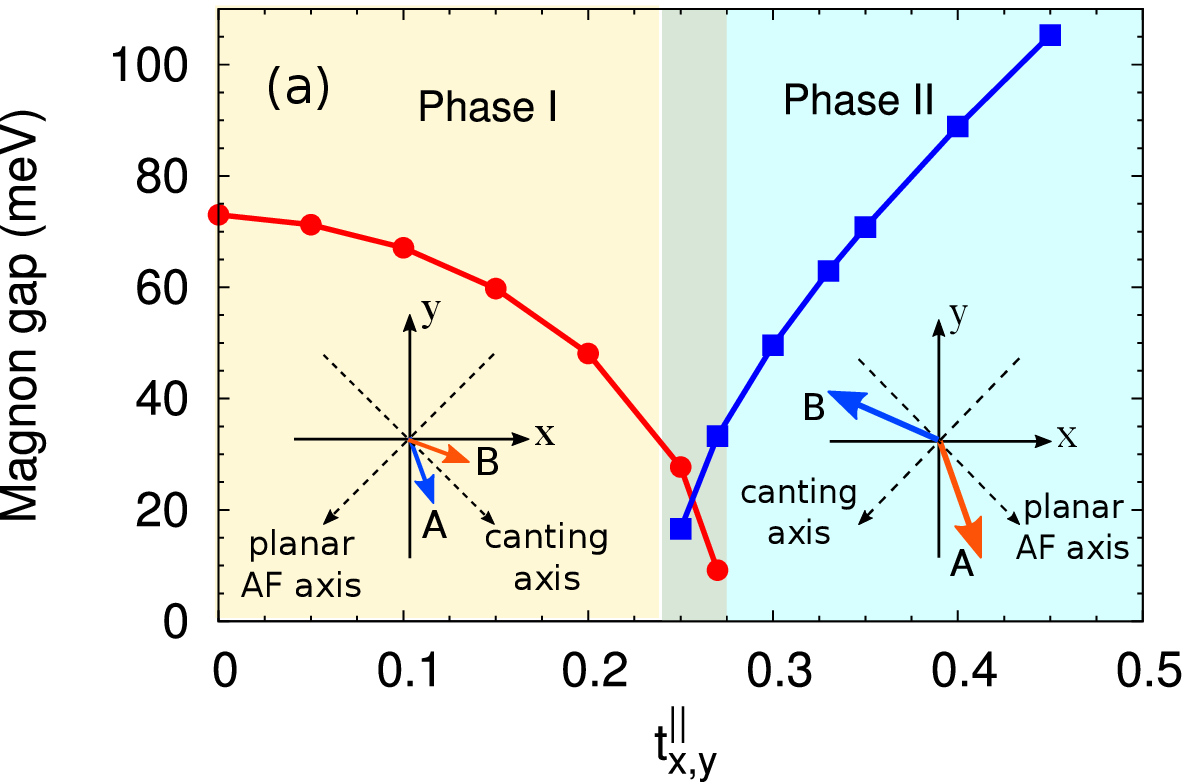,angle=0,width=80mm}
\psfig{figure=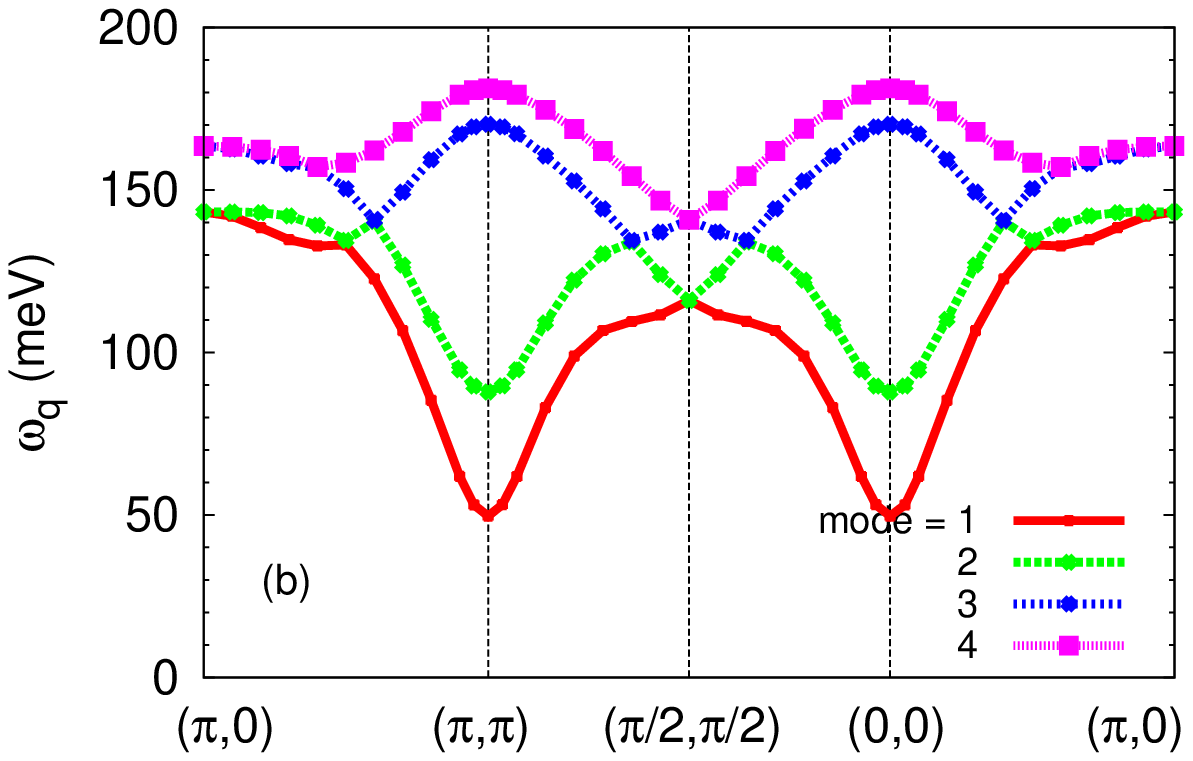,angle=0,width=80mm}
\vspace{0mm}
\caption{(a) Strong magnon gap variation with the octahedral tilting induced hopping terms $t_{x,y} ^\parallel$, showing the AFM I-II reorientation transition. (b) Magnon dispersion in the phase II for $t_{x,y} ^\parallel = 0.3$. Inset in (a) shows only the planar part of the magnetic order. Phases I and II are dominantly $c$-axis and $ab$-plane AFM orders.} 
\label{fig4}
\end{figure}

With increasing $t_x^\parallel=t_y^\parallel$ corresponding to the octahedral tilting, the magnon gap in phase I decreases continuously to zero, as shown in Fig. \ref{fig4}(a). At the critical value ($\sim 0.25$), there is reorientation transition from the dominantly $c$-axis AFM order to the dominantly $ab$-plane AFM order. The hysterisis behaviour near the transition point reflects the divergence in the number of iterations required for self consistency. The magnon gap in phase II increases robustly from zero with $t_{x,y} ^\parallel$ beyond the critical value. Thus, the magnon gap behaviour with $t_{x,y}^\parallel$ is consistent with the AFM I-II reorientation transition as obtained from the self-consistent determination of magnetic order. In phase II, the magnon dispersion shows characteristic differences as seen in Fig. \ref{fig4}(b). Due to mixing between the acoustic and optical modes, the degeneracy is lifted near $(\pi,0)$ and $(\pi/2,\pi/2)$ points. 

Reading off from Fig. \ref{fig4}(a), magnon gap of around 50 meV (as recently reported for the iridate bilayer SL), corresponds to $t_{x,y}^\parallel \approx 0.2$. As this $t_{x,y}^\parallel$ value is quite close to the critical value ($\approx 0.25$), the iridate bilayer SL is on the verge of the reorientation transition. In the following, we summarize the magnon gap trend in the three physical cases considered above. Starting with the bulk case with large octahedral rotation, magnon gap $\approx 90$ meV is obtained, as measured for the bulk compound $\rm Sr_3 Ir_2 O_7$.\cite{sala_PRB_2015} In the reference case (corresponding to reduced octahedral rotation and no tilting), the magnon gap reduces to $\approx 75$ meV ($t_{x,y}^\parallel=0$ in Fig. 4(a)). Finally, in the SL case (with both octahedral rotation and tilting), magnon gap further reduces to $\approx 50$ meV.

\section{Conclusions}

The realistic Hubbard model approach for the bilayer iridate superlattice, with spin-dependent hopping terms directly related to the orbital mixings arising from the octahedral tilting and rotation, provides fundamental insight into the experimentally observed magnon gap reduction associated with the proximity to the isospin reorientation transition from dominantly $c$-axis to $ab$-plane canted AFM order. The spin model analysis shows that the reorientation transition is driven by a reduction in the effective $c$-axis anisotropy term as compared to the bilayer bulk case due to the tilting induced Kitaev terms. Our study indicates the possibility of magnetic order switching by tailoring the octahderal tilting in the bilayer iridate superlattice or via applied pressure in the bilayer bulk compound. \\

\end{document}